\documentclass{iopart}

\usepackage{color,graphicx}
\usepackage{amssymb}
\usepackage{mathrsfs}
\usepackage{supertabular}
\usepackage{color,graphicx}
\usepackage[bookmarksopen]{hyperref}

\def  \f    {\frac}

\def  \del  {\partial}

\def  \bef  {\begin{figure}}
\def  \eef  {\end{figure}}
\def  \be   {\begin{equation}}
\def  \ee   {\end{equation}}
\def  \ba   {\begin{array}}
\def  \ea   {\end{array}}
\def  \bea  {\begin{eqnarray}}
\def  \eea  {\end{eqnarray}}
\def  \beq  {\begin{eqnarray}}
\def  \eeq  {\end{eqnarray}}
\def  \nn   {\nonumber}
\def  \bd   {\begin{displaymath}}
\def  \ed   {\end{displaymath}}
\def  \bse  {\begin{subequations}}
\def  \ese  {\end{subequations}}
\def  \bwt  {\begin{widetext}}
\def  \ewt  {\end{widetext}}

\def  \ba   {{\bf{a_1}}}

\begin{document}

\title[]{Modifications to the pulsar kick velocity due to magnetic interactions in dense plasma}

\author{S P Adhya, P K Roy and A K Dutt-Mazumder}

\address{High Energy Nuclear and Particle Physics Division, Saha Institute
of Nuclear Physics,
1/AF Bidhannagar, Kolkata-700 064, INDIA}
\ead{souvikpriyam.adhya@saha.ac.in}

\begin{abstract}
In this work we calculate pulsar kick velocity of magnetized neutron star composed of degenerate quark matter core with non-Fermi liquid (NFL) correction. Both the leading order (LO) and next to leading order (NLO) corrections to
the kick velocity have been incorporated. In addition, the NFL corrections to the specific heat of magnetized quark matter have been presented. This has been taken into account to calculate the kick velocity of the neutron star. Results show significant departure from the normal Fermi liquid estimates. The relation between radius and temperature has been shown with kick velocity of $100 km/s$ with and without NFL corrections. 

\end{abstract}

\maketitle

\section{Introduction}
 Exploration of the phenomenon of pulsar kicks \rm{i.e.} the observed
large escape velocities of neutron stars (NS) out of supernova remnants has
 drawn significant attention in recent years \cite{spruit98, Brisken, Hobbs, blaschke06}.
Understanding the origin of such high velocities, the so called kick velocities and its connection with pulsar spin has been the subject of many theoretical models \cite{horowitz98, burrows92, burrows95, harrison75, colpi02, Lai, Wang}. 
 However, one of the most natural explanations for these large escape velocities could be due to anisotropic neutrino emission from quark matter present at the core of the NS \cite{chugai84,dorofeev85,vilenkin95,horowitz97} with which we are presently concerned. 
 
 It is well known that neutrinos carry away almost all of the energy released during the supernova and following proto-neutron star evolution. It is estimated that an asymmetry of roughly about $3$ percent in neutrino momentum is sufficient enough to generate such phenomenal velocities \cite{sagert08, sagertarxiv1}.

For a NS with quark matter core in normal phase, the main mechanism of neutrino emission has been the quark direct and inverse URCA processes given by \cite{iwamoto81,shapiro_book},
\bea
&&d\rightarrow u+e^-+\bar{\nu_{e}}
\label{dir}
\eea
\bea
&&u+e^-\rightarrow d+\nu_{e}
\label{inv}
\eea
the above two reactions. These scatterings in presence of a strong magnetic field can give rise to asymmetric neutrino emission as explained in \cite{dorofeev85,sagert08, sagertarxiv1}.

The possibility of the presence of strong magnetic fields in the NS core has been suggested by various authors. At the core of a NS, the magnetic field strength can go upto $\sim 10^{18} G$ as shown in \cite{lai91} using scalar virial theorem. There are few other models too, which support similar values of the magnetic field \cite{yuan98, bandopadhyay97, chakraborty97}. On the surface of the NS, relatively weaker magnetic fields of $\sim 10^{14} G$ have been estimated \cite{woltjer64, mihara90, chanmugam92}. Such high magnetic field could lead to higher asymmetry in the neutrino momentum distribution. This asymmetric neutrino emission is actually related to the electron polarisation opposite to the direction of the magnetic field \cite{chugai84, dorofeev85}.

The electron polarisation for different conditions of magnetic field and kick velocities has been studied recently by Sagert et. al.\cite{sagert08, sagertarxiv1}. In these works, the authors have discussed pulsar acceleration mechanism based on asymmetric neutrino emission from quark direct URCA process in the core of the NS. Here the dependence of the kick velocity with the quark phase temperature and quark phase radius of the star with varying quark chemical potentials has been studied. 

It is to be noted, that for such a high magnetic field strength, the electron motions are quantized. Hence in the present treatment, we need to invoke quantized energy states populated by various Landau levels which shall be discussed later. The critical value for which this quantisation effect becomes important is given by $B_{cric}^e\sim4.4\times10^{13} G$ for electrons and $B_{cric}^q\sim10^2\times B_{cric}^e G$ for quarks respectively. In fact in the strong magnetic field limit what we consider here, electrons are likely to occupy only the lowest Landau level wth electron spin opposite to the direction of magnetic field.

 As the motivation of the present work is to calculate the pulsar kick velocity of magnetized NS composed of degenerate quark matter, we need to know various related quantities like specific heat \cite{holstein73, ipp04} and  neutrino emissivity \cite{schafer04, adhya12} in presence of strong magnetic field. These quantities in turn determine the kick velocity. It might be mentioned in this context that the specific heat of the degenerate quark matter or the emissivity at such high quark density receive significant corrections from the QCD sector which shows NFL behavior in the relativistic regime due to dominance of the magnetic interaction over the electric one. In the non-relativistic limit however, such magnetic corrections due to transverse gluon mediated processes are $(v/c)$ suppressed \cite{manuel00,bellac97}. 

 The central aim of the present work is to expose the role of magnetic interactions on various thermodynamic quantities as well as quark dispersion relations in ultradegenerate quark matter. It is now known from a series of works that in the relativistic regime the magnetic interaction dominates over its electric counterpart \cite{bellac97,manuel00}. For instance, one can mention that for normal FL, the damping width which is actually related to quark-quark scattering rates goes as $\gamma\sim(E-\mu)$ unlike the Coulomb plasma where $\gamma\sim(E-\mu)^2$ \cite{bellac97}. Similar departure have also been observed in the calculation of specific heat capacity and pressure \cite{holstein73,ipp04}. Examples of momentum drag and diffusion coefficients can also be cited where similar deviation from the normal FL behavior has been observed \cite{sarkar10,sarkar11,sarkar13}. In particular, the NFL behavior of the specific heat is most important in the the present context as it directly determines the cooling rate of 
the star. The 
NFL correction in this case involve $T\ln(1/T)$ term which has been dubbed as anomalous corrections in many of the recent literatures \cite{holstein73,rebhan05,manuel00,pal11}. However the entire calculation did not consider the possibility of NFL correction to the kick velocity which we consider in the present work. 
 
 Furthermore, in presence of strong magnetic field, the specific heat will also receive NFL corrections. The latter has been studied extensively for unmagnetized matter but for magnetized matter, so far previous works are confined only to non-interacting quark matter $(\sim g_jqBT/6)$ \cite{xuewen05}. Hence, two major points which we highlight here is the inclusion of NFL correction to the pulsar kick velocity and the calculation of specific heat in presence of strong magnetic field. In our work, leading order (LO) refers to the anomalous logarithmic term 
$T\log (1/T)$ that occurs as the first term in the non-Fermi liquid (NFL) 
correction to the one-loop quark self energy. Quantities such as specific heat
($C_v$) or entropy ($S$) calculated with this term are called LO corrections. In the present work, NLO terms include all other terms
 beyond LO that contain the fractional powers of $T$ and upto $T^3\log (1/T)$ 
that occurs in the expression of the quark self energy. Similarly, quantities 
calculated with this term are labelled as NLO corrections.
 
 The plan of the paper is as follows. In section II we discuss the formalism to calculate the kick velocity and calculate the velocity for different conditions of the magnetic field taking into account the NFL effect in specific heat capacity of magnetized quark matter. We present the results in section III followed by conclusion in section IV.
\section{Formalism}
 Neutron star is composed mainly of neutrons with a fraction of protons. Proton fraction may depend on various models. In presence of magnetic field, the proton fraction may increase \cite{goyal99}. We, however focus on the core of the NS where the density is expected to be so high that the relevant degrees of freedom may be considered to be quarks rather than hadrons \cite{collins75,lattimer01,page00,glen80,haensel94,sagerarxiv0808}. The pulsar kick velocity can be calculated if the luminosity, quark phase radius of the NS and the neutrino emissivity is known. The emissivity is related, as we shall see, with the specific heat of the core.
 Initially, total neutrino luminosity (L) can be approximated as \cite{sagert08, sagertarxiv1},
 \bea
 L\simeq\f{4}{3}\pi R^3 \varepsilon
 \eea
 where R is the quark phase radius of the NS and $\varepsilon$ is the neutrino emissivity.
 Thus, due to momentum conservation we have,
 \bea
 \f{dv}{dt}M_{NS}=\chi L
 \eea
 In the above expression, the polarisation fraction of the electrons has been denoted by $\chi$ and mass of the NS by $M_{NS}$. The total neutrino emission is polarised
in a specific direction. Generally, it is considered that only a fraction $\chi$
of all neutrinos are polarised in one direction. The value of this fraction 
depends on number of spin polarised electrons in comparison to the total number of
electrons.
 Therefore the kick velocity can be written as \cite{sagert08, sagertarxiv1},
\begin{eqnarray}
 dv=\frac{\chi}{M_{NS}}\frac{4}{3}\pi R^{3}\varepsilon dt
 \label{diffv}
\end{eqnarray}
 Using the cooling equation, 
\begin{eqnarray}
 C_{v}dT=-\varepsilon dt
 \label{cooleq}
\end{eqnarray}
one can rewrite Eq.(\ref{diffv}) in terms of the specific heat of the quark matter core.
Eq.(\ref{diffv}) and Eq.(\ref{cooleq}) allow one to calculate the pulsar kick velocity. Recently, such a calculation has already been performed in \cite{sagert08,sagertarxiv1}. However there,  the modification of the specific heat  due to presence of magnetic field has not been considered. We also extend the previous calculation by incorporating LO and NLO correction in the specific heat with and without incorporating the effects due to the presence of magnetic field in the specific heat.
\subsection{Specific heat capacity of degenerate magnetized quark matter}
It will be interesting to see how the thermodynamic potential and associated quantities are modified in the presence of external magnetic field in the core of the NS. In this section, we shall investigate the effect of strong magnetic field responsible for the quantisation of the orbital motion of charged particles already known in literature as Landau diamagnetism.
 Now, in the presence of a constant magnetic field $(B)$ along the $z$ axis, the thermodynamic potential is modified as \cite{strickland12, chakraborty96},
\bea
\Omega^B=-\f{g_dT|q|B}{2\pi^2}\sum_{\nu=0}^{\infty}\int_0^\infty dp_z\log(1+e^{\beta(\mu-\epsilon)})
\eea
where $\epsilon=\sqrt{p_z^2+m^2+2\nu|q|B}$ is the single particle energy eigen value and $g_d$ is the quark degeneracy. For the case of spin-$1/2$ particles the orbital angular momentum $(n)$ is related to $\nu$ as \cite{landaustat1},
\bea
2\nu=2n+s+1
\eea
where $\nu=0,1,2,..$ and $s=\pm1$ refers to spin up $(+)$ or down $(-)$ states.
Now, we use the following identity to evaluate the integration,
\bea
\int_0^\infty \f{h(\epsilon)d\epsilon}{e^{\beta(\epsilon-\mu)}+1}=\int_0^{\mu}h(\epsilon)d\epsilon +\f{\pi^2}{6}T^2h^{'}(\mu)+\f{7\pi^4}{360}T^4h^{'''}(\mu)+...
\eea
Thus, for the case of lowest Landau level $(\nu=0)$, we obtain,
\bea
\Omega^B &=&-\f{g_d|q|B}{2\pi^2}\int_0^\infty \f{\epsilon d\epsilon}{e^{\beta(\epsilon-\mu)}+1}\nn\\
&&\simeq-\f{g_d|q|B}{2\pi^2}\Big[\f{\mu^2}{2}+\f{\pi^2 T^2}{6}\Big]
\eea
Considering $u$ and $d$ quarks only, we get the total heat capacity per unit volume as,
\bea
C_{v}\Big{|}_{FL}^{B}=\f{N_CN_fTm_q^2}{6}\Big(\f{B}{B_{cr}^q}\Big)
\label{cvFLB}
\eea
which can be obtained from,
\bea
S=-\Big(\f{\del\Omega}{\del T}\Big)_\mu
\eea
and the entropy and specific heat of the quarks in the regime of low temperature is given as,
\bea
C_v\simeq T\Big(\f{\del S}{\del T}\Big)_{\mu}
\eea
\begin{figure}[htb]
\begin{center}
\resizebox{8.5cm}{4.75cm}{\includegraphics{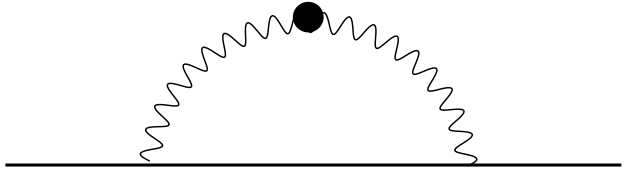}}
\caption{Quark self-energy with resummed gluon propagator.
\label{figQSE}}
\end{center}
\end{figure}
 The QCD interaction was not considered in earlier literatures for the calculation of the specific heat of quarks. Recently, it has been shown by several authors that this interaction mediated by gluons can lead to significant corrections in the specific heat which essentially leads to the NFL behavior \cite{holstein73,ipp04}.
 To extend the calculation of the specific heat beyond the FL result, one needs to incorporate the interaction term in the picture. Mathematically, this requires the knowledge of the one loop quark self energy (QSE).
 To evaluate the one loop quark self energy in degenerate plasma, one 
needs to consider Fig.(\ref{figQSE}), where the solid line represents the fermion 
propagator and the wavy line with the blob implies that the gluon propagator used here is a 
hard dense loop corrected propagator. In the domain of high density and low temperature $(T\sim |\epsilon-\mu|<<g\mu<<\mu)$ which is the region of interest in the context of neutron star the QSE can be written as follows \cite{rebhan05},
 \begin{eqnarray}
\Sigma&\simeq&-g^2C_Fm\,
  \Big\{{(\epsilon-\mu)\over12\pi^2m}\Big[\log\Big({4\sqrt{2}m\over\pi (\epsilon-\mu)}\Big)+1\Big]\nn\\
  &&+{i (\epsilon-\mu)\over24\pi m}
   \,+{2^{1/3}\sqrt{3}\over45\pi^{7/3}}\left({(\epsilon-\mu)\over m}\right)^{5/3}(\mathrm{sgn}(\epsilon-\mu)-\sqrt{3}i)\qquad\nn\\
   &&+ {i\over64 \sqrt{2}}\left({(\epsilon-\mu)\over m}\right)^2
   -20{2^{2/3}\sqrt{3}\over189\pi^{11/3}}\left({(\epsilon-\mu)\over m}\right)^{7/3}(\mathrm{sgn}(\epsilon-\mu)+\sqrt{3}i)\qquad\nn\\
 &&-{6144-256\pi^2+36\pi^4-9\pi^6\over864\pi^6}\Big({(\epsilon-\mu)\over m} 
   \Big)^3 \Big[\log\left({{0.928}\,m\over (\epsilon-\mu)}\right)\nn\\
 &&-{i\pi\mathrm{sgn}(\epsilon-\mu)\over 2}  \Big]
+\mathcal{O}\Big(\left({(\epsilon-\mu)\over m}\right)^{11/3}\Big) \Big\}.
\end{eqnarray}
Now, incorporating the NFL LO result into the expression of the entropy (or specific heat) in absence of any external magnetic field by using the real part of QSE as calculated already in the work of \cite{holstein73, ipp04},
\bea
C_v\Big{|}_{LO}=\f{N_g}{\pi^2}\int_0^{\infty}dp p^2 \f{\del f}{\del T}\f{g_{eff}^2}{24\pi^2}(p-\mu)\log\Big(\f{m^2}{(p-\mu)^2}\Big)
\eea
where $m^2=m_D^2/2$ and the Debye mass ($m_D$) without magnetic field is \cite{rebhan05},
\bea
m_D^2\simeq\f{N_fg^2\mu^2}{2\pi^2}
\eea
In presence of external magnetic field $(B)$, the above equation is modified as,
\bea
C_v^{(i)}\Big{|}_{LO}^B=\Big(\f{g^2C_F}{24\pi^2}\f{|q_i|g_{di}B}{2\pi^2}\Big)\sum_{\nu=0}^{\infty}\int_{0}^{\infty}d\epsilon \f{\del f(\epsilon)}{\del T}(\epsilon-\mu)\log\Big(\f{m_B^2}{(\epsilon-\mu)^2}\Big)
\eea
where i=u or d quark. Here we have assumed the quarks to be ultra-relativistic and confined to $\nu=0$. Substituting $z=\beta(\epsilon-\mu)$ and as $T<<\mu$, we obtain,
\bea
C_v^{(i)}\Big{|}_{LO}^B=\Big(\f{g^2C_F}{24\pi^2}\f{|q_i|g_{di}B}{2\pi^2}T\Big)\int_{-\infty}^{\infty}\f{z^2e^z}{(1+e^z)^2}\log\Big(\f{m_B^2}{T^2z^2}\Big)dz
\eea
Solving the above integration, we obtain at the order $T\log T$,
\bea
C_v\Big{|}_{LO}^B\simeq\Big(\f{N_CN_fC_f\alpha_s}{36\pi}\Big)m_q^2\Big(\f{B}{B_{cr}^q}\Big)T\Big[(-1+2\gamma_E)+2log\Big(\f{2m_B}{T}\Big)\Big]
\eea
Now, the NLO contribution to the specific heat in absence of magnetic field is given as,
\bea
C_v\Big{|}_{NLO}&=&\f{N_g}{\pi^2}\int_0^\infty dp p^2\f{\del f}{\del T}g_{eff}^2m\Big[0.003\left({(\epsilon-\mu)\over m}\right)^{5/3}\nn\\
&&-0.00420\left({(\epsilon-\mu)\over m}\right)^{7/3}\nn\\
&&+0.001\Big({(\epsilon-\mu)\over m}\Big[\log\left({{0.928}\,m\over (\epsilon-\mu)}\right) 
   \Big)^3\Big]
\eea
Similarly, using the NLO terms (second $\log$ term plus the fractional powers of $T$) in the QSE, the NLO contribution to the specific heat capacity in presence of external magnetic field is obtained as,
\bea
C_v\Big{|}_{NLO}^B\simeq&&\Big(\f{N_CN_f}{3}\Big)(C_f\alpha_s)\Big(m_q^2\f{B}{B_{cr}^q}\Big)T\Big[c_1\Big(\f{T}{m_B}\Big)^{2/3}\nn\\
&+&c_2\Big(\f{T}{m_B}\Big)^{4/3}+c_{3}\Big(\f{T}{m_B}\Big)^{2}\Big(c_{4}-\log\Big(\f{T}{m_B}\Big)\Big)\Big]
\eea
where the constants are,
\bea
c_1=-0.2752; c_2=0.2899; c_3=-0.5919; c_4=5.007.
\eea
The Debye mass ($m_B$) in the QCD case in presence of magnetic field is obtained as follows,
\bea
m_B^{2}=\f{N_fg^2m_q^2}{4\pi^2}\Big(\f{B}{B_{cr}^q}\Big)
\eea

We shall also briefly mention the heat capacity in a non-magnetized quark matter for comparison.
It can be recalled that in general, the thermodynamic potential $(\Omega)$ is defined as \cite{landaustat2},
\bea
\Omega=-Tg_d\int \f{d^3p}{(2\pi)^3}\log(1+e^{\beta(\mu-\epsilon)})
\eea 
 For the case of ultrarelativistic degenerate quarks, we obtain,
\bea
\Omega=-\f{N_CN_f\mu^4}{12\pi^2}-\f{N_CN_f\mu^2T^2}{6}
\eea
In the region of low temperature and high chemical potential, the Fermi liquid expression of the specific heat of quark is given as \cite{holstein73},
\begin{eqnarray}
C_{v}\Big|_{FL}&=&\frac{N_{c}N_{f}}{3}\mu_{q}^{2}T
\label{cvFL}
\end{eqnarray}
where $N_C$ and $N_f$ are the number of color and flavor factors respectively. We have taken a 2 flavor system comprising of $u$ and $d$ quarks. 
the effect of $s$ quark is neglected as it is not fully relativistic due to its high mass \cite{iwamoto81,chakraborty96}. 
The specific heat of the degenerate quark matter in absence of external magnetic field up to NLO is given by \cite{holstein73,ipp04},
\bea 
\label{spec-heat}
 \label{finalcv}
 C_v\Big{|}_{total}=C_v\Big{|}_{FL}+C_v\Big{|}_{LO}+C_v\Big{|}_{NLO}
 \eea
 where,
 \bea
 C_v\Big{|}_{LO}=N_g{g_{eff}^2\mu_q^2 T\over36\pi^2}\left(\ln\left({4g_{eff}\mu_q\over\pi^2T}\right)+
   h_0\right)
 \eea
 and
 \bea
 C_v\Big{|}_{NLO}&&=N_g\Big[
  h_1 T^{5/3}(g_{eff}\mu_q)^{4/3}
   +
   h_2 T^{7/3}(g_{eff}\mu_q)^{2/3}\nn\\
   &&+
 h_3 T^3
   \Big[\ln\left({g_{eff}\mu_q\over T}\right)
   +
   h_4\Big]\Big]
 \eea
 where the constants are,
 \bea
h_0=-1.8529; h_1=-0.0402; h_2=0.1089; h_3=-0.6198
 \eea
 and
 \bea
 h_4=3.5159.
 \eea
 The coupling constant $g$ is related to $g_{eff}$ as,
 \begin{equation}\label{geffdef}
 g^2 = \frac{2\ g^{2}_{eff}}{N_{f}},
 \end{equation}
 and $C_v\Big{|}_{total}$ is the sum of the FL, LO and NLO contribution to the specific heat of the quark matter.

\subsection{Kick velocity}
Pulsar kick velocity incorporating the effect of the magnetic field on the specific heat to the best of our knowledge has not been reported earlier.
The pulsar kick velocity obtained taking into account the magnetic field effect on the specific heat capacity of the quarks reads as,
\bea
 v\Big|_{FL}^B &\simeq&\frac{4.15 N_{C}N_f}{3}\Big(\frac{\sqrt{m_q^2(B/B_{cr}^q)}}{400MeV}\frac{T}{1MeV}\Big)^{2}\Big(\frac{R}{10km}\Big)^{3}\frac{1.4M_{\odot}}{M_{NS}}\chi\frac{
km}{s}
\label{vBFL}
\eea
The NFL LO result is obtained as,
\bea
v\Big|_{LO}^B &\simeq&\frac{8.8 N_{C}N_f}{3}(C_f\alpha_s)\Big(\frac{\sqrt{m_q^2(B/B_{cr}^q)}}{400MeV}\frac{T}{1MeV}\Big)^{2}\Big(\frac{R}{10km}\Big)^{3}\nn\\
&&\times\frac{1.4M_{\odot}}{M_{NS}}\Big[0.0635+0.05\log\Big(\f{m_B}{T}\Big)\Big]\chi\frac{
km}{s}
\label{vBLO}
\eea
The NLO correction to the kick velocity can be written as,
\bea
v\Big|_{NLO}^B&\simeq&\frac{8.3 N_{C}N_{f}}{3}\Big(\f{B}{B_{cr}^q}\Big)\Big(\frac{m_{q}}{400MeV}\frac{T}{1MeV}\Big)^{2}\Big(\frac{R}{10km}\Big)^{3}\frac{1.4M_{\odot}}{M_{NS}}\nn\\
&&\times\chi(C_{F}\alpha_{s})\Big[a_1\Big(\frac{T}{m_B}
\Big)^{2/3}+a_2\Big(\frac{T}{m_B}\Big)^{4/3}\nn\\
&&+ \Big[a_3+a_4\ln\Big(\frac
{m_B}{T}\Big)\Big]\Big(\frac{T}{m_B}\Big)^2\Big]\frac{km}{s}
\label{vBNLO}
\eea
The constants are evaluated as,
\begin{eqnarray}
 a_1=-\frac{12\pi\times0.04386}{8};a_2=\frac{12\pi\times0.04613}{10}
;a_3=-2.4162
\end{eqnarray}
and 
\bea
a_4=-0.4595
\eea
where the electron polarisation fraction $(\chi)$ is taken as the case may be. In Eqs. (\ref{vBFL}),(\ref{vBLO}) and (\ref{vBNLO}), $qB$ has been replaced by $m_q^2(B/B_{cr}^q)$.

The net contribution to the pulsar kick velocity upto NLO is obtained by the sum of the
Fermi liquid result and the non-Fermi liquid correction upto NLO:
\bea
v\Big{|}_{total}=v\Big|_{FL}+v\Big|_{LO}+v\Big|_{NLO}
\label{vtotal}
\eea
 The Fermi liquid contribution to the pulsar kick velocity as reported in \cite{sagert08, sagertarxiv1} can be recast into the following form,
 \begin{eqnarray}
  v\Big|_{FL}&\simeq&\frac{8.3 N_{C}N_{f}}{3}\Big(\frac{\mu_{q}}{400MeV}\frac{T}{1MeV}\Big)^{2}\Big(\frac{R}{10km}\Big)^{3}\frac{1.4M_{\odot}}{M_{NS}}\chi\frac{
 km}{s}
 \label{v1FL}
 \end{eqnarray} 
 Thus we obtain the LO contribution to the kick velocity as,
 \begin{eqnarray}
 v\Big|_{LO}&&\simeq\frac{16.6 N_{C}N_{f}}{3}(C_F\alpha_s)\Big(\frac{\mu_{q}}{400MeV}\frac{T}{1MeV}\Big)^{2}\Big(\frac{R}{10km}\Big)^{3}\nn\\
 &&\times\frac{1.4M_{\odot}}{M_{NS}}\chi\Big[c_1+c_2\ln\Big(\f{g\mu_q\sqrt{N_f}}{T}\Big)\Big]\frac{
 km}{s}
 \label{v1LO}
 \end{eqnarray}
 where $C_F=(N_C^2-1)/(2N_C)$ and the constants are $c_{1}=-0.13807$ and $c_{2}=0.0530516$.
 Now we have also extended our calculation beyond the LO in NFL correction. The NLO correction to the pulsar kick velocity is obtained as,
 \begin{eqnarray}
  v\Big|_{NLO}&\simeq&\frac{16.6 N_{C}N_{f}}{3}\Big(\frac{\mu_{q}}{400MeV}\frac{T}{1MeV}\Big)^{2}\Big(\frac{R}{10km}\Big)^{3}\frac{1.4M_{\odot}}{M_{NS}}\chi(C_{F}\alpha_{s})\nn\\
  &\times&\Big[a_1\Big(\frac{bT}{\mu_q}
 \Big)^{2/3}+a_2\Big(\frac{bT}{\mu_q}\Big)^{4/3}\nn\\
 &+& \Big[a_3+a_4\ln\Big(\frac
 {\mu_q}{b T}\Big)\Big]\Big(\frac{bT}{\mu_q}\Big)^2\Big]\frac{km}{s}
 \label{v1NLO}
 \end{eqnarray} 
  and
 \begin{eqnarray}
  b=\frac{2\pi}{\sqrt{N_f}g}.
 \end{eqnarray}
 The long range magnetic interactions lead to an anomalous $T^2\rm{ln}T^{-1}$ term in the expression of the pulsar kick velocity. Eqs.(\ref{v1FL}),(\ref{v1LO}) and (\ref{v1NLO}) will be used to calculate the kick velocity with fully polarised electrons ($\chi=1$) and note that we have neglected the effect of the magnetic field on the specific heat of the quarks. 

In order to evaluate $\chi$ for different cases, we first choose the magnetic field strength to be much larger than the
temperature, the chemical potential as well as the electron mass ($\mu_e, m_e,T
\ll\sqrt{2eB}$). The number density in this case is given by \cite{sagert08, sagertarxiv1},
\begin{eqnarray}
n_\mp=\frac{eB}{(2\pi)^2}\sum_{\eta}\int_0^{\infty}dp\frac{1}{e^{\sqrt{p^2+2\eta
eB}/T}+1},
\end{eqnarray}
The electron polarisation is given as \cite{sagert08, sagertarxiv1},
\begin{eqnarray}
\chi&\sim& 1-\frac{4}{\ln(2)}\sqrt{\frac{\pi T}{2\sqrt{2eB}}}e^{-\sqrt{2eB}/T}.
\label{pol_4}
\end{eqnarray}
where the second term is exponentially small for large magnetic field and one can assume $\chi\sim1$ in this limit. In that case, the kick velocities are approximately same as obtained by using Eqs.(\ref{vBFL}), (\ref{vBLO}), (\ref{vBNLO}) and Eqs.(\ref{v1FL}), (\ref{v1LO}), (\ref{v1NLO}) with $\chi=1$.
The total kick velocity up to NLO is given as the sum of the FL, LO and NLO contribution to the kick velocity.

We next consider the case of weak magnetic field in which case we neglect the effect of magnetic field on the specific heat of quark matter. A comparison between Eq.(\ref{cvFL}) and Eq.(\ref{cvFLB}) shows that for a given value of weak field, the contribution from Eq.(\ref{cvFL}) is dominant. 
In presence of magnetic field ($B$) the number density of electrons is given
by:
\begin{eqnarray}
n_\mp=\frac{geB}{(2\pi)^2}\sum_{\eta} \int_0^{\sqrt{\mu_e^2-m_e^2-2\eta eB}} dp_z.
\end{eqnarray}
The number of Landau levels is limited to $\eta_{max}=\frac{\mu_e^2-m_e^2}{2eB}$.
For the case of $(\mu_e^2-m_e^2)\gg2eB$, the number of occupied Landau levels is large and the sum can be replaced by integration over $\eta$ where $\eta$ is the Landau level number.
Thus we obtain,
 for the case of vanishing
temperature (cold neutron stars), the electron spin polarisation is
\cite{sagert08, sagertarxiv1},
\begin{eqnarray}
 \chi\simeq\frac{3}{2}\frac{m_{e}^2}{\mu_{e}^2-m_{e}^2}\Big(\frac{B}{B_{cr}^e}
\Big)
\label{chilowB}
\end{eqnarray}
where the critical value of the magnetic field is given by
$B_{cr}^e\simeq4.4\times10^{13} G$.
In this case for the sake of brevity we assume that the specific heat does not depend on the external magnetic field.
Thus including the effect of electron spin polarisation on the FL contribution to the kick velocity, Eq.(\ref{v1FL}) is modified as,
\begin{eqnarray}
 v\Big|_{FL}&&\simeq\frac{8.3N_{C}N_{f}}{2}\Big(\frac{\mu_{q}}{400MeV}\frac{T}{1MeV}\Big)^{2}\Big(\frac{R}{10km}\Big)^{3}\frac{1.4M_{\odot}}{M_{NS}}\nn\\
 &&\times\Big(\f{m_e^2}{\mu_e^2-m_e^2}\f{B}{B_{cr}^e}\Big)\frac{
km}{s}
\label{vBwFL}
\end{eqnarray}
The LO contribution to the kick velocity is given by,
\begin{eqnarray}
v\Big|_{LO}&&=\frac{16.6 N_{C}N_{f}}{2}(C_F\alpha_s)\Big(\frac{\mu_{q}}{400MeV}\frac{T}{1MeV}\Big)^{2}\Big(\frac{R}{10km}\Big)^{3}\frac{1.4M_{\odot}}{M_{NS}}\nn\\
&&\times\Big(\f{m_e^2}{\mu_e^2-m_e^2}\f{B}{B_{cr}^e}\Big)\Big[c_1+c_2\ln\Big(\f{g\mu_q\sqrt{N_f}}{T}\Big)\Big]\frac{
km}{s}.
\label{vBwLO}
\end{eqnarray}
We have also extended our calculation beyond the LO in NFL correction. The NLO correction to the pulsar kick velocity turns out to be,
\begin{eqnarray}
 v\Big|_{NLO}&&\simeq\frac{16.6 N_{C}N_{f}}{2}\Big(\frac{\mu_{q}}{400MeV}\frac{T}{1MeV}\Big)^{2}\Big(\frac{R}{10km}\Big)^{3}\frac{1.4M_{\odot}}{M_{NS}}\nn\\
 &&\times\Big(\f{m_e^2}{\mu_e^2-m_e^2}\f{B}{B_{cr}^e}\Big)(C_{F}\alpha_{s})
 \Big[a_1\Big(\frac{bT}{\mu_q}
\Big)^{2/3}+a_2\Big(\frac{bT}{\mu_q}\Big)^{4/3}\nn\\
&&+ \Big[a_3+a_4\ln\Big(\frac
{\mu_q}{b T}\Big)\Big]\Big(\frac{bT}{\mu_q}\Big)^2\Big]\frac{km}{s}.
\label{vBwNLO}
\end{eqnarray}
The total velocity is given by Eq.(\ref{vtotal}).

\section{Results and discussions}
An estimation of the quark phase radius of the NS as a function of temperature (R-T relationship) has been presented in this section for different kick velocities and different conditions of magnetic field in NS. For this purpose, we have assumed the quark chemical potential to be $400 MeV$ with temperature ranging from $3$ to $20$ MeV. Our parameters are in good agreement with the high baryon density prevalent at the core of the NS. In addition, we have taken $\alpha_s=0.5$ and $\mu_e=10$ MeV \cite{sagert08, sagertarxiv1}. The pulsar mass has been taken to be $1.4M_{\odot}$ (where $M_{\odot}=2\times 10^{30} kg$ is the mass of the Sun). As high magnetic fields are persistent in the core of the NS, we have taken magnetic fields of $\sim 10^{15}$G or higher ($\sim 10^{19}$G) relevant to the corresponding cases \cite{ferrer10}. In our work, we have assumed an order of magnitude estimate $(~100 km/s)$ as the initial value of the kick velocity to present realistic values of quark core radius as found in 
NS for the $R$ vs. $T$ plots. These calculations have 
been carried 
out by 
incorporating the effect of the presence and absence of external magnetic field effect on the specific heat of the degenerate quark matter in the core. In addition, the FL, NFL LO and NLO results are given for comparison for all the cases.

\begin{figure}[]
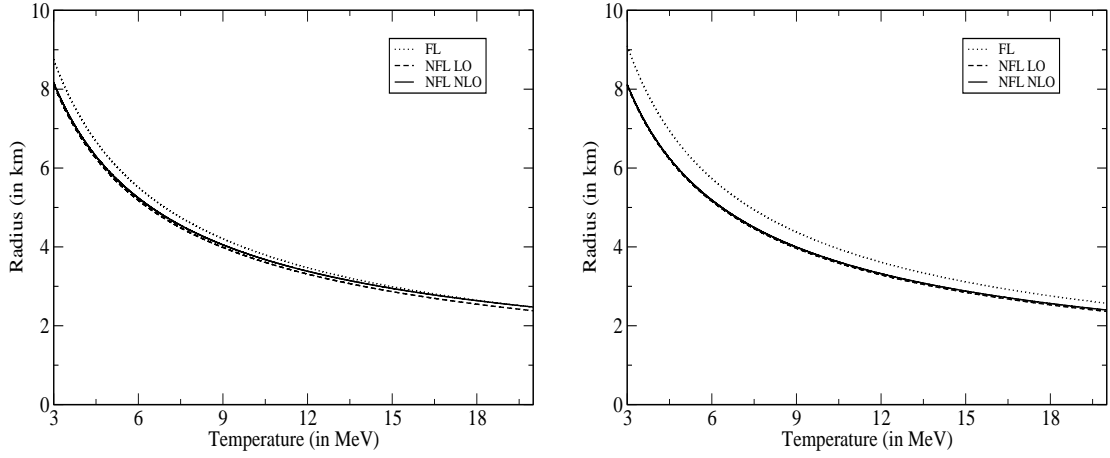

 \bigskip
\bigskip
\begin{center}
 \resizebox{7.0cm}{6.0cm}
{\includegraphics{vkickk1b0v2.eps}}~~~~~~~
 \resizebox{7.0cm}{6.0cm}
{\includegraphics{vkickk1nz.eps}}
\caption{The figure shows the comparison between the FL, NFL LO and NFL NLO result for the radius and temperature dependence for the case of fully polarised electrons. Results have been plotted for the case of presence and absence of external magnetic field effect in the specific heat of degenerate quark matter. The left panel shows the relationship (FL,LO and NLO respectively) for the case where external magnetic field on the specific heat is ignored. The right panel shows the corresponding case when magnetic field $(B=5\times10^{19} G)$ effect in specific heat is included. }
\label{fig1}
\end{center}
\end{figure}

In the left panel of Fig.(\ref{fig1}) we observe that there is a considerable decrease in the quark phase radius of the NS with the inclusion of the LO correction over the FL case. 
For this case, we have considered Eqs.(\ref{v1FL}),(\ref{v1LO}) and (\ref{v1NLO}) (for left panel) for the comparison of the R-T behavior.
This entails a significant increment of the kick velocity for a given temperature due to NFL correction. 
The NFL LO result increases by $21.54 \%$ over the FL result for a radius and temperature of $5$ km and $5$ MeV for the quark core when external magnetic field in specific heat is not included. In addition, the change is $38.54 \%$ on including the magnetic field effect in specific heat. Similarly, the NFL NLO decreases by $2.84 \%$ and $1.28 \%$ over the NFL LO when magnetic field effect is excluded and included respectively.
We have obtained the result considering fully polarised electrons, thus making the electron polarisation fraction ($\chi$) equal to unity. We have also plotted the behavior of the the R-T relationship (right panel) for the kick velocity where the magnetic field effect is taken in the specific heat and for this we use Eqs.(\ref{vBFL}),(\ref{vBLO}) and (\ref{vBNLO}). We note that there is a considerable decrease in the radius of the quark phase due to the anomalous NFL LO corrections. However there is only a modest dimunition in the radius of the quark phase of the NS due to NFL NLO correction to the kick velocity for all the cases.

\begin{figure}[]
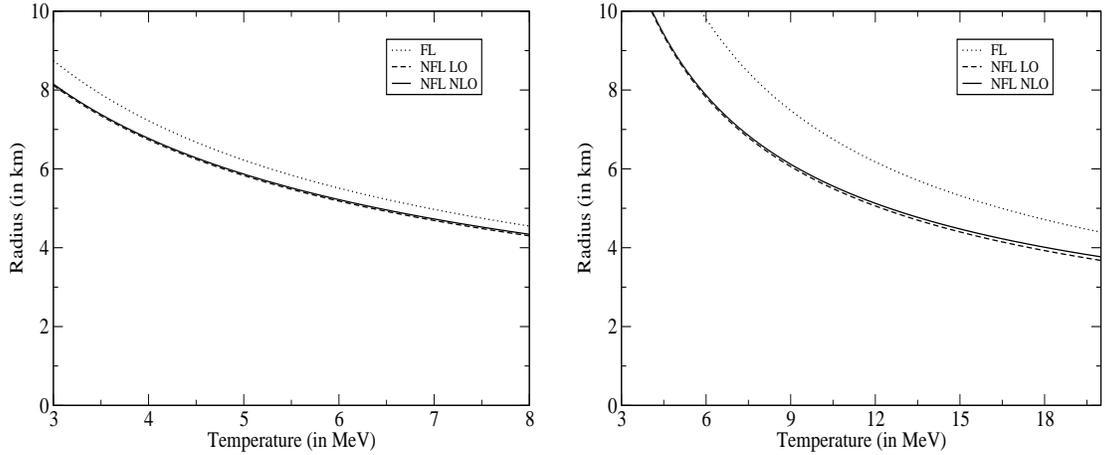

\bigskip
 \bigskip
\begin{center}
\resizebox{7.0cm}{6.0cm}{\includegraphics{kickvekkihighcv0.eps}}~~~~~~~\resizebox{7.0cm}{6.0cm}{\includegraphics{kickvelkihighcvB.eps}}
\caption{The figure shows the numerical comparison where high magnetic field has been taken into account along with vanishing temperature for kick velocity of $100 km/s$. The left panel shows the relationship (FL,LO and NLO respectively) for the case where external magnetic field on the specific heat is ignored. The right panel shows the corresponding case when magnetic field $(B=10^{19} G)$ effect in specific heat is included. }
\label{fig2}
\end{center}
\end{figure}

 In Fig.(\ref{fig2}), for a large value of magnetic field, we have generated the plot numerically for the case of highly polarised electrons as in Eqn.(\ref{pol_4}).
 We have found that there is an appreciable change of the $R-T$ relationship of the NS due to LO although the results are not very different from the fully polarised case for a given order. However, the NLO correction to the kick velocity does not show considerable change from the LO result as evident from the graphs. In the left panel, we have ignored the effect of the magnetic field on the specific heat whereas in the right panel we have considered the effect of the external magnetic field on the specific heat.
 
\begin{figure}[]
\begin{center}
\resizebox{7.0cm}{6.0cm}{\includegraphics{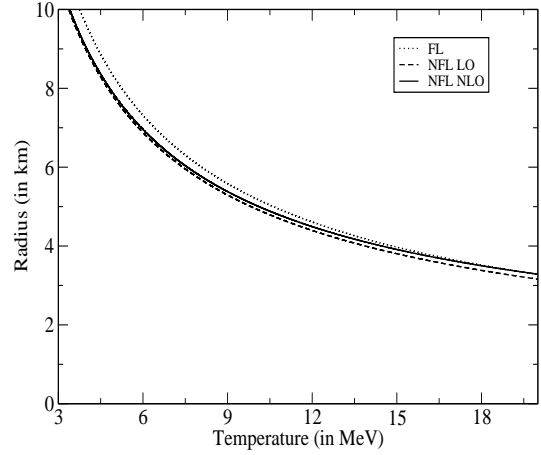}}
\caption{The figure shows the comparison between the FL, NFL LO and NFL NLO result for the radius and temperature dependence for the case of partially polarized electrons in weak magnetic field $(B=5\times10^{15} G)$ for kick velocity of $100 km/s$. Results have been plotted for the case of absence of external magnetic field effect in the specific heat of degenerate quark matter. }
\label{fig3}
\end{center}
\end{figure}

\begin{figure}[]
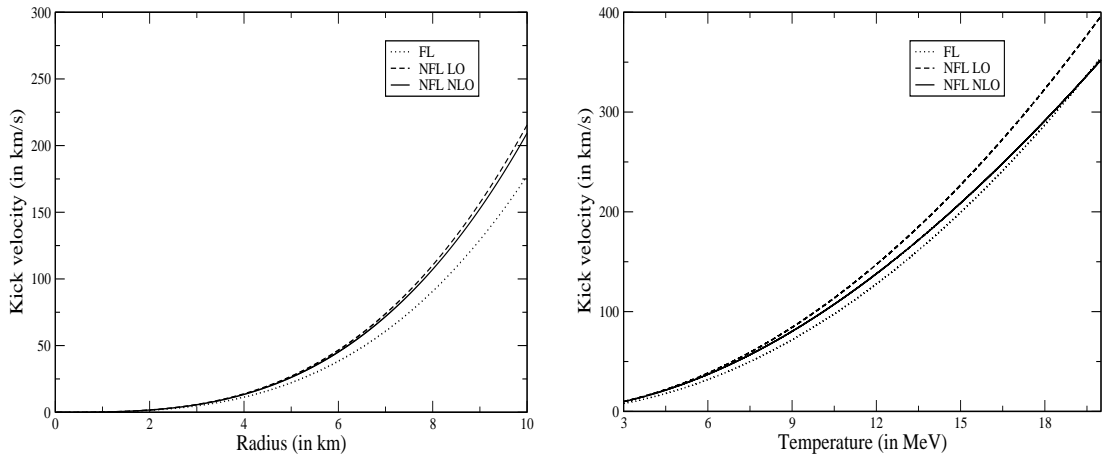

\begin{center}
\resizebox{7.0cm}{6.0cm}{\includegraphics{vkickvsRkless.eps}}~~~~~~~\resizebox{7.0cm}{6.0cm}{\includegraphics{vkickvsTkless.eps}}
\caption{The figure shows the comparison between the FL, NFL LO and NFL NLO result for the kick velocity dependence with the radius (left panel) and temperature (right panel) for the case of partially polarized electrons in weak magnetic field $(B=5\times10^{15} G)$. Results have been plotted for the case of absence of external magnetic field effect in the specific heat of degenerate quark matter. A temperature of $5 MeV$ has been assumed for the left panel and a quark core radius of $5 km$ has been assumed for the right panel. }
\label{fig4}
\end{center}
\end{figure}

   Fig.(\ref{fig3}) shows the comparison between the FL, LO and NLO result for the radius and temperature dependence for the case of partially polarized electrons for kick velocities of $100 km/s$.
  Equations (\ref{vBwFL}),(\ref{vBwLO}) and (\ref{vBwNLO}) are plotted with temperature as the independent variable and radius of the quark phase as dependent variable. The LO correction lowers the quark phase radius of the NS for a specific value of temperature. We have found out that the NLO results impose a slight correction to the LO results. In this case, we have ignored the effect of the magnetic field on the specific heat of the quarks which is insignificant in the weak field limit.
  In Fig.(\ref{fig4}), using Eqs. (\ref{vBwFL}),(\ref{vBwLO}) and (\ref{vBwNLO}), we have shown the behavior of the kick velocity with the radius (left panel) and temperature (right panel). A typical value of temperature of $5$ MeV and a quark core radius of $5$ km have been assumed for left and right panel respectively.
  In both the panels of Fig.(\ref{fig4}), we note there is a considerable increase in the kick velocity of NS due to NFL LO corrections over the FL result. But, the NFL NLO correction reduces the kick velocity as compared to the NFL LO result.

\section{Conclusion}
  In this work, we have derived the expressions of the pulsar kick velocity including the NFL corrections to the specific heat of the degenerate quark matter core. The contributions from the electron polarization ($\chi$) for different cases has also been taken into account to calculate the velocities. In addition, comparison has been made between the NFL LO and NLO contributions to the kick velocity with the FL case.
  We have included the effect of the external magnetic field into the specific heat of the degenerate quark matter for the calculation of the pulsar kick velocity. The calculation of the specific heat of the degenerate quark matter in magnetic field for the NFL LO and NLO are new and have not been reported earlier. For all the remaing cases, a kick velocity of the order of $\sim 10^2 km/s$ is generated which is still low for a wide class of pulsars having average speed of about $\sim 10^3 km/s$.  We have found that the NFL LO contributions are significant while calculating the radius-temperature relationship as seen from 
the graphs presented for the case of the neutron star with moderate and high magnetic field. The anomalous corrections introduced to the pulsar kick velocity due to the NFL (LO) behavior increases appreciably the kick velocity for a particular value of radius and temperature. However, for all the cases, no appreciable change in the R-T relationship has been observed for the NLO correction with respect to the LO case. It is to be noted that we have performed our calculations for magnetic field as high as $10^{19}$G. Such high magnetic field may not be realised in the core of a neutron star. Moreover, for magnetic fields larger than $10^{18}$G, the longitudinal pressure becomes negative resulting in mechanical instability of quark matter \cite{isayev13}. As magnetic field enhances the effect of the NFL correction to the kick velocity, we have assumed such high value for the estimation of the maximum velocity due to anisotropic neutrino emission from the quark core taking into account NFL corrections to 
the specific heat.

Finally, we note that we have considered 2-flavour system as $s$ quark is
not fully relativistic. However, it might be mentioned that both direct
and inverse Urca processes for $s$ quark becomes important beyond certain
critical density. In addition, introduction of $s$ quark may influence
the neutrino emissivity by modifying the population of electrons
and their spin polarization. The effect of including the strange quarks
into the system should be investigated separately. 
\section{Acknowledgments}
One of the authors [SPA] would like to thank UGC, India for providing the fellowship (Sr. no.: 2120951147) during the tenure of this work. SPA would also like to thank S. Sarkar for useful discussions regarding different aspects of this paper.


\section*{References}
\begin{thebibliography}{50}
\bibitem{spruit98} Spruit H and Phinney E S 1998 {\it Nature} {\bf393} 139.
\bibitem{Brisken}  Chatterjee S, Vlemmings W H T, Brisken W F et~al 2005 {\it APJL}
  {\bf 630} L61.
\bibitem{Hobbs}
Hobbs G, Lorimer D, Lyne A and Kramer M 2005 {\it MNRAS} {\bf 360} 974.
\bibitem{blaschke06} Berdermann J, Blaschke D, Grigorian H, Voskresensky D N 2006 {\it Prog. in Part. and Nucl. Phys.} {\bf 57} 334-342.
\bibitem{horowitz98} Horowitz C J and Li G 1998 {\it Phys. Rev. Lett.} {\bf 80} 3694.
\bibitem{burrows92}  Burrows A and Fryxell B A 1992 {\it Science} {\bf 258} 430.
\bibitem{burrows95} Burrows A, Hayes J and Fryxell B A 1995 {\it Ap. J.} {\bf 450} 830.
\bibitem{harrison75} Harrison E R and Tademaru E 1975 {\it Ap. J.} {\bf 201} 447.
\bibitem{colpi02} Colpi M and Wasserman I 2002 {\it Ap. J.} {\bf 581} 1271.
\bibitem{Lai}
 Lai D, Chernoff D F \&  Cordes J M 2001 {\it APJ} {\bf 549} 1111.
\bibitem{Wang}
 Wang C, Lai D, \& Han J L 2006 {\it APJ} {\bf 639} 1007.
 \bibitem{chugai84}  Chugai N N 1984 {\it Sov. Astron. Lett.} {\bf 10} 87.
\bibitem{dorofeev85} Dorofeev O F, Radionov V N and Ternov I M 1985 {\it Soviet
Astronomy Letters} {\bf 11} 123.
\bibitem{vilenkin95}  Vilenkin A 1995 {\it Ap. J.} {\bf 451} 700.
\bibitem{horowitz97}  Horowitz C J and Piekarewicz J 1997 {\it arXiv 9701214}.
\bibitem{sagert08} Sagert I and Schaffner-Bielich J 2008 {\it J. Phys. G} {\bf 35} 014062.
\bibitem{sagertarxiv1} Sagert I and Schaffner-Bielich J {\it arXiv 0708.2352}.
\bibitem{iwamoto81}
 Iwamoto N 1981 {\it Ann. Phys. (N.Y.)} {\bf141} 1.
\bibitem{shapiro_book} Shapiro S L and Teukolsky S A 1983 {\it Black Holes, 
White Dwarfs and Neutron Stars} Wiley-Interscience New York.
\bibitem{lai91} Lai D and Shapiro S L 1991 {\it Astrophys. J.} {\bf 383} 745.
\bibitem{yuan98} Yuan Y F and Zhang J L 1998 {\it A. and A.} {\bf 335} 969-972.
\bibitem{bandopadhyay97} Bandopadhyay D, Chakraborty S and Pal S 1997 {\it Phys. Rev. Lett.} {\bf 79} 12.
\bibitem{chakraborty97}  Chakraborty S, Bandopadhyay D and Pal S 1997 {\it Phys. Rev. Lett.} {\bf 78} 2898.
\bibitem{woltjer64} Woltjer C 1964 {\it Ap. J.} {\bf 140} 1309.
\bibitem{mihara90} Mihara T A et. al. 1990 {\it Nature (London)} {\bf 346} 250.
\bibitem{chanmugam92} Chanmugam G 1992 {\it Annu. Rev. Astron. Astrophys.} {\bf 30} 143.
\bibitem{holstein73} Holstein T, Norton R E and Pincus P 1973 {\it Phys. Rev. B} {\bf
8} 2649.
\bibitem{ipp04} Gerhold A, Ipp A and Rebhan A 2004 {\it Phys.Rev.D} {\bf 70} 105015
; {\bf 69} R011901.
\bibitem{schafer04} Sch\"{a}fer T and Schwenzer K 2004 {\it Phys.Rev.D} {\bf 70}
114037.
\bibitem{adhya12} Adhya S P, Roy P K and Dutt-Mazumder A K 2012 {\it Phys.Rev.D} {\bf 86} 034012.
\bibitem{manuel00} Manuel C 2000 {\it Phys.Rev.D} {\bf 62} 076009.
\bibitem{bellac97} Le Bellac M and Manuel C 1997 {\it Phys.Rev.D} {\bf 55} 3215.
\bibitem{sarkar10} Sarkar S and Dutt-Mazumder A K 2010 {\it Phys.Rev.D} {\bf 82}
056003.
\bibitem{sarkar11} Sarkar S and Dutt-Mazumder A K 2011 {\it Phys.Rev.D} {\bf 84}
096009.
\bibitem{sarkar13} Sarkar S and Dutt-Mazumder A K 2013 {\it arXiv 1209.5153}.
\bibitem{rebhan05} Gerhold A and Rebhan A 2005 {\it Phys.Rev.D} {\bf 71} 085010.
\bibitem{pal11} Pal K and Dutt-Mazumder A K 2011 {\it Phys.Rev.D} {\bf 84} 034004.
\bibitem{xuewen05} Xuewen L, Xiaoping Z and Defu H 2005 {\it Astroparticle Phys.} {\bf 24} 92-99.
\bibitem{goyal99} Goyal A 2009 {\it Phys.Rev.D} {\bf 59} 
101301(R).
\bibitem{collins75} Collins J C and Perry M J 1975 {\it Phys. Rev. Lett.} {\bf 34} 1353.
\bibitem{lattimer01} Lattimer J M and Prakash M 2001 {\it Astrophys. J.} {\bf 550} 426.
\bibitem{page00} Page D, Prakash M, Lattimer J M and Steiner A W 2000 {\it Phys. Rev. Lett.} {\bf 85} 2048.
\bibitem{glen80} Glen G and Sutherland P 1980 {\it Ap. J.} {\bf 239} 671.
\bibitem{haensel94} Haensel P and Gnedin O Y 1994 {\it A. \& A.} {\bf 290} 458.
\bibitem{sagerarxiv0808} Sagert I, Pagliara G, Hempel M and Schaffner-Bielich J {\it arXiv 0808.1049v1}.
\bibitem{strickland12} Strickland M, Dexheimer V and Menezes D P 2012 {\it Phys.Rev.D} {\bf 86} 125032.
\bibitem{chakraborty96} Chakraborty S 1996 {\it Phys.Rev.D} {\bf 54} 1306.
\bibitem{landaustat1} Landau L D and Lifshitz E M {\it Statistical Physics (Pergamon Press New York)} Vol 5 Part I.
\bibitem{landaustat2} Landau L D and Pitaevskii L P {\it Statistical Physics (Pergamon Press New York)} Vol 9 Part II.
\bibitem{ferrer10} Ferrer E J, de la Incera V, Keith J P, Portillo I and Springsteen P L 2010 {\it Phys.Rev.C} {\bf 82} 065802.
\bibitem{isayev13} Isayev A A and Yang J 2013 {\it J. Phys. G} {\bf 40} 035105.
\end{thebibliography}
\end{document}